\input{aipcheck}
\documentclass[sort&compress
    ,final            
  ]
  {aipproc}

\layoutstyle{6x9}

\newcommand{\beq}{\begin{equation}}
\newcommand{\eeq}{\end{equation}}
\newcommand{\half}{\frac 1 2}
\usepackage{amsmath}
\usepackage{amsfonts}
\usepackage{amssymb}
\begin{document}

\title{An overview of f(R) theories}


\author{Santiago Esteban Perez Bergliaffa}{address={Departamento de F\'{\i}sica Te\'orica, Instituto de Física, Universidade do Estado do Rio de Janeiro, Brazil}
}

\begin{abstract}
A brief introduction to theories of the gravitational field with a Lagrangian that is a function of 
the scalar curvature is given. The emphasis will be placed in formal developments, while 
comparison to observation will be discussed in the chapter by S. Jor\'as in this volume.

\end{abstract}

\maketitle

\section{Introduction}

The predictions of General Relativity (GR) are confirmed to an impressive degree by observations in number of situations
\cite{will}. In spite of this fact, theories that differ from GR either in the limit of low or high curvature have 
been intensively studied lately, and have 
a long tradition, starting with a paper by Weyl in 1918 \cite{weyl}. Although Weyl's motivation was related to the unification of GR and Electrodynamics, the current revival of these theories is twofold. 
In the case of low curvature, the aim is to describe the accelerated expansion of the universe that follows from several observations \cite{chinos}
(when interpreted in the standard cosmological model \cite{inhom})
\footnote{The possibility of describing the current accelerated expansion of the 
universe using $f(R)$ theories was first discussed in 
\cite{capoz}.}. Regarding the high-curvature regime, it is important to 
note that
there is no observational evidence of the behaviour
of the gravitational field
for very large values of the curvature. This makes objects such as black holes and neutron stars the ideal places to 
look for deviations from General Relativity in the strong regime.
In fact, the Kerr solution is not unique in $f(R)$
theories \cite{psaltis}. Consequently, any deviation from Kerr's spacetime in compact objects will be unequivocally
signaling the need for changes in our description of strong gravity.
The task of understanding what kind of deviations can be expected, and their relation to observable quantities
is of relevance
in view of several developments 
that 
offer the prospect of observing properties 
of black holes in the vicinity of the horizon
\cite{bhobs}.

In this short review we will be concerned with gravitational theories 
described by the action
\beq
S = \int d^4x \sqrt{-g} f(R),
\eeq
where $g$ is the determinant of the metric $g_{\mu\nu}$ and $f$ is an arbitrary function of the curvature scalar $R$
\footnote{This choice is favoured by a theorem by Ostrogradski \cite{ostro} over Lagrangians built  with invariants obtained from the the Ricci and Riemann tensors.}.
The function $f$ must satisfy certain constraints, some of which are necessary for the theory to be well-defined
{\textit{ab initio}},
and others to account for observational facts. Those in the first class will be discussed in this review, while 
those in the second class are presented in the chapter by S. Jor\'as in this volume
\footnote{There are several reviews that deal with different aspects of $f(R)$ theories, see \cite{reviews},\cite{noodre},\cite{reviewdft},\cite{sotifara}, and also the recent book \cite{capofara}.}.
We shall begin by reviewing in the next section
some general features of 
this type of theory.

\section{The three versions of $f(R)$ theories}

We shall see in this section that $f(R)$ theories can be 
classified in three different types, according to the role attibuted to the connection.
In all of the versions, the equation for the energy-momentum conservation is valid, 
since the total (gravitational plus matter) action is diffeomorphism-invariant and gravity and matter are minimally coupled by hypothesis (see for instance \cite{wald}, \cite{koi}). 

\subsection{Metric version}

In the metric version of $f(R)$ theories, the action
\beq
S=\frac{1}{2\kappa}\int d^4x \sqrt{-g}f(R)+S_{\rm M}(g_{\mu\nu},\psi),
\label{actf}
\eeq
is varied with respect to $g_{\mu\nu}$. Here, $S_{\rm M}$ is the matter action, which is independent of the connection.
The resultant equations of motion are of fourth order in the derivatives of the metric tensor:
\beq
\frac{df(R)}{dR}R_{\mu\nu}-\half f(R)g_{\mu\nu}-\left[\nabla_\mu\nabla_\nu-g_{\mu\nu}\Box\right]\frac{df(R)}{dR}=
\kappa T_{\mu\nu},
\label{action}
\eeq
where $T_{\mu\nu}$ is the energy-momentum of he matter fields, defined by
$$
T_{\mu\nu} = -\frac{2}{\sqrt{-g}}\frac{\delta S_{\rm M}}{\delta g^{\mu\nu}},
$$
and the covariant derivative 
is defined using the usual Levi-Civita connection. 
Taking the trace, we obtain
$$
\frac{df(R)}{dR}R-2f(R)+3\Box\frac{df(R)}{dR}=\kappa T,
$$
which is to be compared to $R=-\kappa T$, the result in GR. 

\subsubsection{Equivalence with Brans-Dicke theory}

As shown for instance in \cite{hindawi}, the gravitational part of the action given in Eqn.(\ref{actf}) is equivalent to
the following action:
\beq
S=\int d^4 x\sqrt{-g}\left[\frac{\phi R}{2\kappa}-U(\phi)\right],
\label{actbd}
\eeq
where 
\beq
U(\phi)=\frac{\phi\chi(\phi)-f(\chi(\phi))}{2\kappa},
\label{pot}
\eeq
$\phi = f_{,\chi}(\chi )$, and $\chi = R$, 
corresponding to a Brans-Dicke theory with $\omega = 0$
\cite{bdbook}.
Note that the absence of a kinetic term for the scalar field does not mean that it is non-dynamical: its evolution, due to the non-minimal coupling with 
the gravitational field, is given by the variation of the action wrt $\phi$:
\beq
3\Box \phi + 2U(\phi ) - \phi \frac{dU}{d\phi} = \kappa T.
\eeq
Through a conformal transformation of the metric and a redefinition of the scalar field,
the action given in Eqn.(\ref{actbd}) can be taken to that of a scalar field minimally coupled with gravity, and with nonzero
kinetic term and potential. 
These representations of $f(R)$ theories 
show that there is a massive scalar degree of freedom, which manifests as a longitudinal mode in   
gravitational radiation (see for instance \cite{gravrad} for the cosmological case).

We close this section by stating that the equivalent representations  
are 
convenient since the associated equations of motion are of order two, but 
a word of caution is needed because sometimes the potential in Eqn.(\ref{pot})
is typically multivalued (see for
instance \cite{joras}\cite{critic}). 
It may be better to work directly in the original representation, as  
for instance in Ref.\cite{salgado} in the case of compact stars. 

\subsection{Palatini version}

In this second type of $f(R)$ theories, the metric and the connection are taken as independent fields, and
the matter action $S_{\rm M}$ is independent
of the connection. 
So the starting point is the action
\beq
S=\frac{1}{2\kappa}\int d^4x\sqrt{-g}\;f({\cal R}) + S_{\rm M}(g_{\mu\nu},\psi),
\eeq
where ${\cal R} = g^{\mu\nu}{\cal R}_{\mu\nu}$, and the corresponding Riemann tensor is constructed
with a connection $\Gamma$ a priori independent of the metric. 

From the variation of the action wrt the metric and $\Gamma$ 
we get \footnote{In the case of GR, this method
furnishes the same result as the metric case, but this is not the case in more general theories 
as discussed for example in \cite{buch}.}
(see for instance \cite{sotifara})
\beq
f'({\cal R}){\cal R}_{(\mu\nu)}- \half f({\cal R}) g_{\mu\nu}= \kappa T_{\mu\nu},
\label{palgr}
\eeq
\beq
\bar\nabla_\lambda (\sqrt{-g}\;f'({\cal R}) g^{\mu\nu})=0, 
\label{palc}
\eeq
where the prime denotes derivative wrt ${\cal R}$, 
and the barred covariant derivative is built with the connection $\Gamma$. 
GR is recovered by setting $f({\cal R}) = {\cal R}$ in these equations. 
Taking the trace of Eqn.(\ref{palgr}) we obtain 
\beq
f'({\cal R}){\cal R}-2f({\cal R})=\kappa T,
\label{alg}
\eeq
which shows that in this case the relation between ${\cal R}$ and $T$ is algebraic, 
hence no scalar mode is present. 

From Eqn.(\ref{palc}), it follows that \cite{sotifara}
\beq
\Gamma^\lambda_{\mu\nu} = \frac{1}{f'({\cal R})}g^{\lambda\sigma}\left[\partial_\mu(f'({\cal R})g_{\nu\sigma})
+\partial_\nu(f'({\cal R})g_{\mu\sigma})-\partial_\sigma(f'({\cal R})g_{\mu\nu})\right].
\eeq
Since this expression relates $\Gamma$ to ${\cal R}$ and the metric, and ${\cal R}$ and $T$ are in principle interchangeable through Eqn.(\ref{alg}), the connection can be in principle expressed in terms of the matter fields and the metric. In other words, it is an auxiliary field. In fact, Eqn.(\ref{palgr}) can be rewritten as
\begin{eqnarray}
G_{\mu\nu} & = & \frac{\kappa}{f'}T_{\mu\nu}-\half g_{\mu\nu}\left({\cal R}-\frac{f}{f'}\right)+
\frac{1}{f'}(\nabla_\mu\nabla_\nu-g_{\mu\nu}\Box)f'
\\ \nonumber
 & & -\frac 3 2 \frac{1}{f'^2}\left[(\nabla_\mu f')(\nabla_\nu f')
-\half g_{\mu\nu}(\nabla f')^2\right]
\label{palmod}
\end{eqnarray}
where the Einstein tensor and the covariant derivatives are built with the Levi-Civita connection, and ${\cal R}$ is expressed in terms of $T$ using Eqn.(\ref{alg}). It follows that this version of $f(R)$ theories can be interpreted as GR with a modified source. Perhaps the most important modification is that third order derivatives of the matter fields  appear on the rhs of Eqn.(\ref{palmod}). As reported in \cite{palstar}, this feature may cause serious problems in static spherically symmetric solutions with a polytropic fluid with index $3/2 < \gamma < 2$ as a source. Note however that this result was challenged in the review \cite{olmo} \footnote{It has also been claimed that the Cauchy problem is not 
well-posed for the Palatini version of $f(R)$ theories \cite{sotifara}. For an updated discussion see \cite{capocau}.}.

\subsection{Metric-affine theories}

In this case, the matter action depends of the connection, which is a priori independent of the metric. The action is given by
\beq
S = \frac{1}{2\kappa}\int d^4x\sqrt{-g}f({\cal R})+S_{\rm M}(g_{\mu\nu},\Gamma^\lambda_{\mu\nu},\psi).
\eeq
Depending on the matter fields, the theory may display non-propagating torsion and non-metricity
(see \cite{sotifara} and \cite{sotili} for details).

\section{Nonminimal coupling}

Metric $f(R)$ theories 
have been generalized by allowing a nonminimal coupling between the curvature and the matter Lagrangian, with action 
given by
\beq
S = \int\left\{\half f_1(R)+[1+\lambda f_2(R)]{\cal L}_{\rm m}\right\}\sqrt{-g}d^4x,
\eeq
where $f_1$ and $f_2$ are arbitrary functions of $R$, and $\lambda$ is a constant. A particular case of this action was considered in 
\cite{nood} in the context of the accelerated expansion of the universe. Later, it was shown in \cite{berto}
\footnote{See also \cite{koi}.} that this type of theory leads to a modification of the conservation law of the matter energy-momentum tensor, namely
\beq
\nabla^\mu T_{\mu\nu}^{({\rm m})} = \frac{\lambda}{1+\lambda f_2} f_2'\left[g_{\mu\nu}{\cal L}_{\rm m}-T_{\mu\nu}^{({\rm m})}\right]\nabla^\mu R.
\eeq
The presence of a nonzero rhs leads to non-geodesic motion, and it was suggested in \cite{berto} that this may be related to MOND. 

A more general type of theories was proposed in \cite{h1}, with action given by 
\beq
S = \int f(R,{\cal L}_{\rm m})\sqrt{-g}d^4x,
\eeq
where $f$ is an arbitrary function of $R$ and of the matter Lagrangian. As in the previous case, an extra force, perpendicular to the 4-velocity, accelerates the particles. 
To close this section, let me mention that all the versions of $f(R)$ theories mentioned above can be described in a unified framework in the so-called C-theory \cite{koivisto}.

\section{Assorted applications}

$f(R)$ theories have been used to describe different aspects of relativistic astrophysics and cosmology. 
Since the low curvature limit, which has been studied primarily to explain the accelerated expansion of the universe, is discussed in the chapter by S. Jor\'as in this volume, only one example will be given here in this regime. Afterwards, some applications in the strong-curvature regime will be discussed.  

\subsection{Low curvature}

In the case of the $k=0$ Friedmann-Lem\^aitre-Robertson-Walker metric, the EOM (\ref{action}) 
can be written as 
\beq
\rho=-f' R_{tt}-\frac{f}{2}+3f''\frac{\dot a \dot R}{a},
\label{rho}
\eeq
\beq
p=-\frac{f'}{3}\left(R_{tt}+R\right)+\frac f 2 -f'' \left( \ddot
R-\frac{2\dot a \dot R}{a}\right)-f'''\dot R^2.
\label{p}
\eeq
Let us remark that it is safe to asssume that most of the current matter content of
the universe
(assumed here to be normal matter, as opposed to dark energy) is pressureless. This matter must satisfy the conditions $\rho_0\geq 0$ and $p_0=0$, where the subindex 0 means that the quantity is evaluated today.
Using Eqns.(\ref{rho}) and (\ref{p}),
we shall rewrite these conditions in terms of following kinematical parameters:
the Hubble and deceleration parameters, the jerk, and the snap,
respectively given by \cite{mattcqg}
$$
H=\frac{\dot a}{a},\,\,\,\,\,\,\,\,\,\,\,\,\,\,\,q=-\frac{1}{H^2}\frac{\ddot a}{a},
\,\,\,\,\,\,\,\,\,\,\,\,\,\,\,
j=\frac{1}{H^3}\frac{\stackrel{...}{a}}{a},\,\,\,\,\,\,\,\,\,\,\,\,\,\,\,s=\frac{1}{H^4}\frac{\stackrel{....}{a}}{a}.
$$
While the current value of the first two parameters is relatively well-established today, 
the value of $j_0$ is not known with high precision, and no acceptable 
value of $s_0$ has been reported yet \cite{vita}. 
By writing
$\rho_0\geq 0$ in terms of the kinematical parameters we get
\beq
3q_0H_0^2f_0'-\frac{f_0}{2} -18 H_0^4f_0''(j_0-q_0-2)\geq 0.
\label{r1}
\eeq
This inequality gives a relation that the parameters and the derivatives of a given $f(R)$ must satisfy today and, as shown in \cite{mine}, it limits the possible values of the parameters of a given theory.
Notice that Eqn.(\ref{p}) involves the snap (through $\ddot R$). If we had a measurement
of $s_0$, we could use the equation
$p_0=0$ to obtain another constraint on $f(R)$. Since this is not the case, we shall express $p_0=0$ in such a way
that it gives
a forecast fo the possible current values of the snap
for a given $f(R)$:
\begin{equation}
s_0  =  \frac{f_0'}{6H_0^2f_0''}(q_0-2)+
 6H_0^2\frac{f_0'''}{f_0''}(-q_0+j_0-2)^ 2-  
[q_0(q_0+6)+2(1+j_0)] -\frac{f_0}{12H^4f_0''}.
\label{r2}
\end{equation}

\subsection{Strong curvature}

\begin{itemize}

\item The possibility of nonsingular cosmological solutions in $f(R)$ theories has been considered in 
\cite{bounce1} and \cite{bounce2}. In the latter article, a
necessary condition for a bounce to occur in a Friedmann-Lemâtre-Robertson-Walker setting 
was obtained, and it is given by
\beq
\frac{\ddot a_0}{a_0} = -\frac{\rho_0}{f'_0}+\frac{f_b}{2f'_0},
\eeq
with
\beq
R_0=6\left(\frac{\ddot a_0}{a_0}+\frac{K}{a_0^2}\right),
\eeq
and the subindex $b$ means that the quantity is evaluated at the bounce. 
Contrary to the case of GR, a bounce may occur for any value of $K$.

\item It was shown in \cite{staro} that the theory given by 
\beq
f(R) = R + \frac{R^2}{(6M)^2}
\eeq
has an inflationary solution given by
$$
H\approx H_i-\frac{M^2}{6}(t-t_i),
$$
$$
a\approx a_i\exp\left[H_i(t-t_i)-\frac{M^2}{12}(t-t_i)^2\right],
$$
where $t_i$ marks the beginning of the inflationary epoch. 
Several features of this model have been studied in detail (see references in
\cite{reviewdft}). The results of WMAP constraint $M\approx 10^{13}$ GeV, and the spectral 
index for this model is $n_{\cal R}\approx 0.964$, which is in the range allowed by WMAP 5-year constraint. 
The tensor to scalar ratio $r$ also satisfies the current observational bound, but is different
from that of the chaotic inflation model. Hence, future observations such as the Planck satellite 
may be able to discriminate between these two models.

\item Compact stars have been repeatedly studied for a number of $f(R)$ theories, either in the conformal
representation \cite{babilan}, or directly in the fourth-order version (see for instance 
\cite{salgado}).

\item Regarding black holes in $f(R)$ theories, it was shown in \cite{psaltis} that 
an observational verification of the Kerr solution for an astrophysical object cannot be used in distinguishing between GR and $f(R)$ theories. Hence, the observation of deviations from the Kerr
spacetime may point to changes
in our understanding of gravitation. Other features of black holes in $f(R)$ theories have been analyzed in
\cite{othersbh}.

\end{itemize}

\section{Conclusions}

In this short review, I intended to show that several aspects of $f(R)$ theories (in its various  
representations) have been extensively studied in the literature. There are many other aspects 
that I had to leave aside such as ``good propagation'' (\textit{i.e.} absence of shocks) \cite{sari}, 
the loop representation \cite{loop}, and the Hamiltonian representation \cite{derru}.
Although the low-curvature regime
and its consequences has attracted 
a lot of attention due to its possible relevance in Cosmology, 
the high-curvature regime is also of interest independently of the low-curvature regime, and the consequences of a modification in such a regime may be testable in the near future
(as in the inflationary model in \cite{staro}, and through the 
observation of electromagnetic \cite{bhobs} and gravitational waves  
\cite{fla}
in the
case of compact objects).

\begin{theacknowledgments}
The author acknowledges support from CNPQ, FAPERJ, UERJ, and ICRANet-Pescara.
\end{theacknowledgments}

\bibliographystyle{aipproc}

\begin{thebibliography}{9}

\bibitem{will} \textit{The Confrontation between general relativity and experiment},
Clifford M. Will, Living Rev.Rel. 9 (2005) 3, \texttt{gr-qc/0510072}.
See in particular 
\textit{Gravity Probe B: Final results of a space experiment to test general relativity},
C. W. F. Everitt et al., Phys. Rev. Lett. 106, 221101 (2011) for recent results of 
Gravity Probe B. See also
I. Ciufolini et al., in \textit{General Relativity and John Archibald
Wheeler}, edited by I. Ciufolini and R. A. Matzner (Springer,
Dordrecht, 2010), p. 371, and \textit{Some considerations on the present-day results for the detection of frame-dragging after the final outcome of GP-B},
Lorenzo Iorio, Europhys.Lett.96:30001 (2011).

\bibitem{weyl} \textit{Gravitation und Elektrizit\"at},  H. Weyl, Sitzungsber. Preuss. Akad. d.
Wiss. Teil 1 (1918) 465-480.

\bibitem{chinos} See for instance \textit{Dark Energy},
Miao Li, Xiao-Dong Li, Shuang Wang, \texttt{1103.5870 [astro-ph.CO]}.

\bibitem{inhom} See for instance 
\textit{Structures in the Universe by Exact Methods - Formation, Evolution, Interactions},         K. Bolejko, A. Krasinski, C. Hellaby and M-N. Célérier, 
Cambridge University Press (2009).

\bibitem{capoz} \textit{Curvature quintessence}, S. Capozziello,
Int. J. Mod. Phys. D11 (2002) 483-492,
\texttt{gr-qc/0201033}.

\bibitem{reviews} \textit{Fourth order gravity: equations, history, and applications to cosmology},
H.-J. Schmidt,
Int. J. Geom. Meth. Mod. Phys.4:209-248,2007, \texttt{gr-qc/0602017}.

\bibitem{noodre} \textit{Introduction to Modified Gravity and Gravitational Alternative for Dark Energy},
S. Nojiri, S.D. Odintsov, Int. J. Geom. Meth. Mod. Phys. 4:115-146 (2007),
Unified cosmic history in modified gravity: from F(R) theory to Lorentz non-invariant models,
Shin'ichi Nojiri, Sergei D. Odintsov, Phys. Rept. 505:59-144 (2011).

\bibitem{reviewdft}
\textit{f(R) theories},
Antonio De Felice, Shinji Tsujikawa,
Living Rev. Rel. 13: 3 (2010),
\texttt{1002.4928}.

\bibitem{sotifara} \textit{f(R) Theories Of Gravity},
Thomas P. Sotiriou, Valerio Faraoni, 
Rev. Mod. Phys. 82, 451-497 (2010),
\texttt{arXiv:0805.1726}.

\bibitem{capofara} \textit{Beyond Einstein Gravity: A Survey of Gravitational Theories for Cosmology and Astrophysics},
Salvatore Capozziello and Valerio Faraoni, Springer (2010).

\bibitem{wald} \textit{General Relativity}, Robert Wald, University of Chicago Press (1984). 

\bibitem{bhobs} See for instance
\textit{Probes and Tests of Strong-Field Gravity with Observations in the Electromagnetic Spectrum}
Dimitrios Psaltis, 
Living Rev. Rel. 11, no. 9 (2008).

\bibitem{ostro} M. Ostrogradski, Mem. Ac. St. Petersbourg VI 4, 385 (1850),
see also \textit{Avoiding dark energy with 1/r modifications of gravity},
R. P. Woodard, Lect. Notes Phys. 720 (2007) 403, \texttt{astro-ph/0601672}.


\bibitem{hindawi} \textit{Nontrivial vacua in higher derivative gravitation},
Ahmed Hindawi, Burt A. Ovrut, Daniel Waldram, Phys. Rev. D53 (1996) 5597, 
\texttt{hep-th/9509147}.

\bibitem{salgado} 	
\textit{Robust approach to f(R) gravity},
L. G. Jaime, L. Patino, M. Salgado, Phys. Rev. D83 (2011) 024039,
\texttt{1006.5747 [gr-qc]}.

\bibitem{bdbook} \textit{The Scalar-Tensor Theory of Gravitation},
Y. Fujii and Kei-ichi Maeda, Cambridge Monographs on Mathematical Physics (2007).

\bibitem{joras} \textit{Viable Singularity-Free f(R) Gravity Without a Cosmological Constant},
V. Miranda, S. E. Jor\'as, I.Waga, M. Quartin, Phys. Rev.
Lett. 102, 221101 (2009).

\bibitem{critic} \textit{Power-law cosmic expansion in f(R) gravity models},
N. Goheer, J. Larena, and P. K. S. Dunsby, Phys. Rev.
D 80, 061301(R) (2009),
\textit{The Future evolution and finite-time singularities in F(R)-gravity unifying the inflation and cosmic acceleration}, S. Nojiri, and S. D. Odintsov, Phys. Rev. D 78, 046006
(2008).

\bibitem{multis} See for instance \textit{Auxiliary fields representation for modified gravity models},
Davi C. Rodrigues, Filipe de O.Salles, Ilya L. Shapiro, Alexei A. Starobinsky, Phys. Rev. D83 (2011) 084028,
\texttt{1101.5028 [gr-qc]}.

\bibitem{buch} \textit{Non-linear Lagrangians and cosmological theory}, H. A. Buchdahl, Mon. Not. Roy. Astron. Soc. {\bf 150}, 1 (1970).

\bibitem{amen} \textit{Conditions for the cosmological viability of f(R) dark energy models},
L. Amendola, R. Gannouji, D. Polarski, S. Tsujikawa, Phys. Rev. D75 (2007) 083504, 
\texttt{gr-qc/0612180}.

\bibitem{koi} \textit{Covariant conservation of energy momentum in modified gravities},
T. Koivisto, Class. Quant. Grav. 23 (2006) 4289,
\texttt{gr-qc/0505128}.

\bibitem{ghost} \textit{Can the dark energy equation-of-state parameter $w$ be less than -1?}, 
S. M. Carroll, M. Hoffman and M. Trodden, Phys. Rev. D 68, 023509 (2003),
\textit{The Phantom menaced: Constraints on low-energy effective ghosts},
J. M. Cline, S. Jeon and G. D. Moore, Phys. Rev. D 70, 043543 (2004).

\bibitem{gravrad} \textit{The Evolution of cosmological gravitational waves in f(R) gravity},
K. N. Ananda, S. Carloni, P.K.S. Dunsby, Phys. Rev. D77 (2008) 024033,
\texttt{0708.2258 [gr-qc]}, \textit{Higher-order gravity and the cosmological background of gravitational waves},
S. Capozziello, M. De Laurentis, M. Francaviglia, Astropart. Phys. 29 (2008) 125, 
\texttt{0712.2980 [gr-qc]}.

\bibitem{palstar} \textit{A No-go theorem for polytropic spheres in Palatini f(R) gravity},
E Barausse, T. P. Sotiriou, J. C. Miller, Class. Quant. Grav. 25 (2008) 062001 
\texttt{gr-qc/0703132 [GR-QC]}.

\bibitem{olmo} \textit{Palatini Approach to Modified Gravity: f(R) Theories and Beyond},
G. J. Olmo, Int. J. Mod. Phys. D20 (2011) 413, \texttt{1101.3864 [gr-qc]}.

\bibitem{capocau} \textit{The Cauchy problem for f(R)-gravity: An Overview},
S. Capozziello, S. Vignolo, \texttt{1103.2302 [gr-qc]}.

\bibitem{nood} \textit{Gravity assisted dark energy dominance and cosmic acceleration},
Shin'ichi Nojiri, Sergei D. Odintsov,
Phys. Lett. B599 (2004) 137, \texttt{astro-ph/0403622}.

\bibitem{berto} \textit{Extra force in f(R) modified theories of gravity},
O. Bertolami, C. G. Boehmer, T. Harko, F. S. N. Lobo, Phys. Rev. D75 (2007) 104016,
\texttt{0704.1733 [gr-qc]}.

\bibitem{h1} \textit{f(R,Lm) gravity},
T. Harko, F. S. N. Lobo, Eur.Phys.J. C70 (2010) 373,
\texttt{1008.4193 [gr-qc]}.


\bibitem{bounce1} \textit{The stability of general relativistic cosmological theory}, J. D. Barrow and A. C. Ottewill, J. of Phys. A16 (1983).

\bibitem{bounce2} \textit{Bounce conditions in f(R) cosmologies},
S. Carloni, P. K. S. Dunsby, D. M. Solomons,
Class.Quant.Grav. 23 (2006) 1913,
\texttt{gr-qc/0510130}.

\bibitem{staro} \textit{A new type of isotropic cosmological models without singularity}, A. A. Starobinsky, Phys.Lett. B 91, 99 (1980).


\bibitem{psaltis} \textit{Kerr Black Holes are Not Unique to General Relativity},
D. Psaltis, D. Perrodin, K. R. Dienes, I. Mocioiu,
Phys. Rev. Lett. 100 (2008) 091101,
\texttt{0710.4564 [astro-ph]}.



\bibitem{othersbh} \textit{A New framework for studying spherically symmetric static solutions in f(R) gravity},
A. M. Nzioki, S. Carloni, R. Goswami, P. K. S. Dunsby,
Phys.Rev. D81 (2010) 084028,
\texttt{0908.3333 [gr-qc]},\textit{Static Spherically Symmetric Solutions in F(R) Gravity},
L. Sebastiani, S. Zerbini,
Eur.Phys.J. C71 (2011) 1591,
\texttt{1012.5230 [gr-qc]}, \textit{f(R) black holes},
T. Moon, Yun Soo Myung, E. J. Son,
\texttt{1101.1153 [gr-qc]}, \textit{Static and spherically symmetric solutions in $f(R)$ theories}, S. E. Perez Bergliaffa and Y. Cifarelli, Phys. Rev. D 84, 084006 (2011).

\bibitem{sotili} \textit{Metric-affine f(R) theories of gravity}, Sotiriou, T. P., and S. Liberati, 2007b, Annals Phys. 322, 935.

\bibitem{babilan} See for instance \textit{Relativistic stars in f(R) gravity},
E. Babichev, D. Langlois, Phys.Rev. D80 (2009) 121501,
\texttt{0904.1382 [gr-qc]}.

\bibitem{sari} \textit{Shock free wave propagation in gauge theories},
J.G. McCarthy, O. Sarioglu, Int. J. Theor. Phys. 39 (2000) 159, 
\texttt{math-ph/9902004}.

\bibitem{loop} \textit{Extension of loop quantum gravity to $f(R)$ theories},
Xiangdong Zhang, Yongge Ma, 
Phys. Rev. Lett. 106, 171301 (2011),
\texttt{1101.1752v2 [gr-qc]}.

\bibitem{derru} \textit{Various Hamiltonian formulations of f(R) gravity and their canonical relationships},
N. Deruelle, Y. Sendouda, A. Youssef,
Phys.Rev. D80, 084032 (2009),
\texttt{0906.4983v1 [gr-qc]}.

\bibitem{fla} See for instance
\textit{The basics of gravitational wave theory}, 
E. Flanagan and S. Hughes, 
New J. Phys., 7, 204, (2005), \texttt{gr-qc/0501041}.

\bibitem{koivisto} \textit{Unifying Einstein and Palatini gravities}
Luca Amendola, Kari Enqvist, Tomi Koivisto,
Phys.Rev.D83:044016 (2011).

\bibitem{mattcqg} See for instance \textit{Jerk and the cosmological equation of state},
M. Visser, Class.Quant.Grav. 21 (2004) 2603, \texttt{gr-qc/0309109}.

\bibitem{vita} See for instance \textit{High-Redshift Cosmography},
V. Vitagliano, Jun-Qing Xia, S. Liberati, M. Viel, JCAP 1003 (2010) 005,
\texttt{0911.1249 [astro-ph.CO]}.

\bibitem{mine} \textit{Constraining f(R) theories with the energy conditions},
S. E. Perez Bergliaffa, Phys.Lett. B642 (2006) 311, 
\texttt{gr-qc/0608072}.

\clearpage
\end{thebibliography}

\end{document}